\documentclass[twocolumn,english,prd,superscriptaddress,nofootinbib,notitlepage]{revtex4-1}

\usepackage[latin1]{inputenc}
\usepackage{babel}

\usepackage{ifpdf}
\ifpdf   
    \usepackage[pdftex]{graphicx}
    \usepackage[hyperfootnotes=false,pdfstartview=,colorlinks=true,linkcolor=purple,citecolor=blue,urlcolor=blue,bookmarks=false]{hyperref}    \DeclareGraphicsExtensions{.png,.pdf}
    \usepackage{pgf}
\else    
    \usepackage[usenames,dvipsnames]{color}
    \usepackage[hyperfootnotes=false,pdfstartview=,colorlinks=true,linkcolor=BrickRed,citecolor=blue,urlcolor=blue,bookmarks=false]{hyperref}
    \usepackage{graphicx}
    \DeclareGraphicsExtensions{.eps}   
\fi

\usepackage{amssymb,amsmath}

\newcommand{\beq}{\begin{eqnarray}}
\newcommand{\eeq}{\end{eqnarray}}
\newcommand{\be}{\begin{eqnarray}}
\newcommand{\ee}{\end{eqnarray}}

\newcommand{\Om}{\Omega}

\newcommand{\La}{\Lambda}

\newcommand{\dd}{\textrm{d}}
\newcommand{\lcdm}{$\Lambda$CDM}

\begin{document}

\title{On the Possibility of Anisotropic Curvature in Cosmology}

\author{Tomi S. Koivisto}
\affiliation{Institute for Theoretical Physics and Spinoza Institute, Utrecht University, 3508 Utrecht, Netherlands}

\author{David F. Mota}
\affiliation{Institute of Theoretical Astrophysics University of Oslo, 0315 Oslo, Norway}

\author{Miguel Quartin}
\affiliation{Institut für Theoretische Physik, Universität Heidelberg, Philosophenweg
16, 69120 Heidelberg, Germany}
\affiliation{Instituto de Física, Universidade Federal do Rio de Janeiro, CEP 21941-972, Rio de Janeiro, RJ, Brazil}

\author{Tom G. Zlosnik}
\affiliation{Perimeter Institute for Theoretical Physics, 31 Caroline Street North, Waterloo, Ontario N2L 2Y5, Canada}

\date{June 16, 2010}

\begin{abstract}
In addition to shear and vorticity a homogeneous background may also exhibit anisotropic curvature. Here a class of spacetimes is shown to exist where the anisotropy is solely of the latter type, and the shear-free condition is supported by a canonical, massless 2-form field. Such spacetimes possess a preferred direction in the sky and at the same time a CMB which is isotropic at the background level. A distortion of the luminosity distances is derived and used to test the model against the CMB and supernovae (using the Union catalog), and it is concluded that the latter exhibit a higher-than-expected dependence on angular position. It is shown that future surveys could detect a possible preferred direction by observing $\sim 20 / (\Omega_{k0}{}^2)$ supernovae over the whole sky.
\end{abstract}

\maketitle


\section{Introduction}
Observational cosmology is undergoing a fast evolution, with the increase of available data and a better understanding of the connections between models and observations. The once ``quasi-sacred'' principles of homogeneity and isotropy are being questioned by several observations of large scale structures~\cite{Copi:2010na}.  This is remarkable, since the Cosmic Microwave Radiation (CMB) is widely regarded a strong evidence for the  $\La$CDM  model which is firmly based on these two principles (sometimes extended to three by an extra assumption of flatness). The $\La$CDM model is, however, baroque, requiring the introduction at different scales and epochs of three sources of energy density: the inflaton, dark matter and dark energy.  This alone should encourage us to continuously challenge the model and the three pillars it stands upon.

In this work, we focus on  a model that, in spite of presenting an
isotropic expansion of the universe, disposes of the other fundamental
pillar of $\La$CDM: it exhibits an anisotropic spatial curvature.  This is realized simply by a 2-form field. While many studies have imposed bounds on the possible presence of universal {\it rotation}~\cite{1985MNRAS.213..917B} and of {\it shear}~\cite{PhysRevLett.77.2883,Graham:2010}, the cosmological and astrophysical implications of solely {\it anisotropic curvature} have not been worked out (see however~\cite{Carneiro:2001fz,salucci,barrow,barrow2,jetp,jetp2,demi,BlancoPillado:2010uw} and references therein for some pioneering work).

We consider scenarios where this anisotropy becomes important at late times. This seems, at first sight, to require  additional fine-tunings. However, the present horizon scale could be linked to the anisotropies by various recent ideas in the literature: in a stringy framework one may naturally find an anisotropic universe tunneling into existence just about 60 e-folds before the end of inflation, corresponding to the scales of the present universe~\cite{Graham:2010,Carroll:2009dn,Adamek:2010}. On the other hand, if the curvature and anisotropy have a common physical origin with the present acceleration of the universe, no new coincidences but rather an observational window on the possible properties of dark energy emerges~\cite{Koivisto:2007b,Koivisto:2008ig,Mota}. In both cases, due to the presence of a 2-form field, one may avoid the usual isotropization theorems: in dynamical compactifications one needs fluxes for stabilization, and on the other hand such may behave as generalized dark energy field~\cite{Koivisto:2009sd}.

We consider homogeneous locally rotationally symmetric (LRS) class of metrics in the context of the idea proposed in~\cite{Mimoso:1993ym} and further developed in~\cite{Coley:1994,McManus:1994}. For simplicity, we choose here the axisymmetric LRS class given by:
\begin{align}\label{eq:lrs-metric}
    \!\!\!\dd s^2 = - \dd t^2 \!+\! a^2(t) \dd y^2 \!+\! b^2(t) \!\! \left[ \dd \xi^2 \!+\! \frac{1}{|k|}S^2(|k|^{\frac{1}{2}}\xi) \dd \phi^2 \right]\!,\!\!
\end{align}
with $S(x) \equiv \{\sin (x), \, x, \,\sinh (x)\}\,$ for $\,\{ k>0,
\,k=0, \linebreak[3] \,k<0\}$, and $k$ is related to the spatial Ricci
scalar via $$^{3}R=\frac{2k}{b^{2}}.$$

For $\,S(x) = \sin (x)\,$ we have the Kantowski-Sachs metric, for
which spatial sections take the form of the product ${\cal R} \times
{\cal S}^{2}$. In this case, as in all cases, the line ${\cal R}$ is
parameterized by the coordinate $y$. However, particular to the this
case, the functions $\xi$ and $\phi$ should be thought of as angular
coordinates on the two sphere ${\cal S}^{2}$.

For $\,S(x) = x\,$ we have a Bianchi I metric with spatial sections of
the form ${\cal R} \times {\cal R}^{2}$. In this case $\xi$ and $\phi$
should be thought of as cylindrical coordinates on the plane ${\cal R}^{2}$. This case reduces to a Friedmann-Robertson-Walker~(FRW)
geometry when $a(t)=b(t)$.

Finally the case $\,S(x)=\sinh(x)\,$ corresponds to a Bianchi III geometry, for which spatial sections take the form of the product ${\cal R}\times{\cal H}^{2}$. Here $\xi$ and $\phi$ should be thought of as hyperbolic coordinates covering the two-hyperboloid ${\cal H}^{2}$.

Using this LRS class of metrics, the authors in~\cite{Mimoso:1993ym} found that if one has an imperfect fluid permeating the universe such that its anisotropic stress $\pi_{ab}$ is related to the electric part $E_{ab}$ of the Weyl tensor associated with the fluid flow by
\begin{equation}
    \label{eq:pi-ab:E-ab}
    \pi_{ab}\;=\;2\,E_{ab}\,,
\end{equation}
then one can have a shear-free anisotropic model the scale factor of which evolves exactly like one in a curved FRW model. This in turn makes both redshift and the Hubble parameter become isotropic.

Although the comoving distance
\begin{equation}\label{eq:comov-dist}
    d_{C}(z) \,\equiv\, d_{\rm comoving}(z) \,\equiv\, \chi \,=\, c \int^z_0 \frac{\dd z'}{H(z')}
\end{equation}
becomes isotropic, as will be shown both the angular diameter and luminosity distances remain anisotropic. It is therefore possible to detect these curvature-induced effects by such distance relations using, say, supernovae (SNe) and the CMB. However, as will be shown, at the background level the CMB is isotropic.

This is a unique feature of these LRS models. Also, since expansion is isotropic one expects to first approximation that the Cosmic Parallax (see~\cite{Quercellini:2009a,Quartin:2010,Quercellini:2009b}) should vanish. Such a model, where the anisotropy is sourced by a scalar field, was investigated in~\cite{Carneiro:2001fz}.


\section{A model for shear-free anisotropic curvature}

To realize this scenario, we follow~\cite{Barrow:1996gx} and consider a canonical two-form field $B_{ab}$. We shall find that this field can indeed act as an imperfect fluid which satisfies~(\ref{eq:pi-ab:E-ab}).

The action for the field is (henceforth $8\pi G \equiv 1$):
\begin{equation}
    S_{B} = \alpha \int J_{abc}J^{abc}\sqrt{-g}d^{4}x \,,
\end{equation}
where $J_{abc}\equiv 3!\nabla_{[a}B_{bc]}$.  There are well-motivated
candidates for such a field in physics, such as the Kalb-Ramond
field~\cite{Kalb:1974yc}.

The equation of motion for  $B_{ab}$ is:
\begin{equation}
    \label{beq2}
    \partial_{a}(\sqrt{-g}J^{abc})=0 \,.
\end{equation}
The resulting stress energy tensor is given by:
\begin{equation}
    \label{stresstb}
    T^{B}_{mn} = -3\alpha J_{mbc}J_{n}^{\phantom{n}bc}+\frac{1}{2}\alpha g_{mn} J_{abc}J^{abc}\,.
\end{equation}

We now make the ansatz that $J_{abc} = f(t)\epsilon_{adbc}V^{d}$ where
$\epsilon_{abcd}$ is the volume 4-form and $V^{d}$ a unit vector
field. We shall see that the field $V^{d}$ defines a `preferred'
direction in spacetime.

For this ansatz the tensor  (\ref{stresstb}) takes the following form
for the LRS geometries described by (\ref{eq:lrs-metric}):
\begin{equation}
\label{tbcosmo}
    T^{B}_{ab} = \rho_{B} U_{a} U_{b} + P_{B} h_{ab}+ L_{B} V_{a}V_{b}\,,
\end{equation}
where
\begin{align*}
    \rho_{B} = \frac{\alpha}{2}J_{abc}J^{abc}\,, \quad
    P_{B} =  -\frac{\alpha}{2}J_{abc}J^{abc}\,,\quad\quad\quad\quad\\
    L_{B} = \alpha J_{abc}J^{abc} \,,\quad
    U^{a} = (\partial_{t})^{a} \,,\quad
    h_{ab} =  g_{ab}+U_{a}U_{b} \,.
\end{align*}

We will concentrate on the case where $V^{a}$  lies along the $y$
direction. The Einstein equations are then (in units where $8\pi G=1$):
\begin{align}
    2\frac{\dot{a}}{a}\frac{\dot{b}}{b}+\frac{k+\dot{b}^{2}}{b^{2}} &=
    \rho+\rho_{B}\,,  \quad
    \frac{\ddot{a}}{a}+\frac{\ddot{b}}{b}+\frac{\dot{a}}{a}\frac{\dot{b}}{b}
    = -P-P_{B}  \nonumber \\
    2\frac{\ddot{b}}{b}+\frac{k+\dot{b}^{2}}{b^{2}}  &= -P - P_{B}- L_{B}\,, \label{ein2}
\end{align}
where $\rho$ and $P$ represent the total contribution to density and pressure from the other fields in the universe.

The only independent components of $J_{abc}$ are $\,J_{t\xi\phi}=
f(t)b^{2}S(\xi)\,$ and permutations thereof. The time dependence of the function $f(t)$ follows from ensuring the satisfaction of (\ref{beq2}) which reduces to:
\begin{align}
    \partial_{t}(f(t)a(t)) = 0
\end{align}
and so we have that $f(t) = C/a(t)$, where $C$ is a constant. It
follows then that
$$J_{abc}J^{abc}=-6 C^{2}/a^{2}\,.$$

From the Einstein equations  (\ref{ein2}) we find the condition $a(t)=b(t)$  (and equivalently the condition (\ref{eq:pi-ab:E-ab})) arises providing that
\begin{align}
    \frac{k}{a^{2}} = -L_{B} =  -\alpha J_{abc}J^{abc}= 6\alpha \frac{C^{2}}{a^{2}}\,.
\end{align}

This equality must arise from a combination of the size of a number appearing in the action (the number $\alpha$) and a number which is reflective of the size $J_{abc}$, and thus of initial conditions on the field (through the constant
$C$). Therefore we have:
$$ \rho_{B} = -k/(2a^{2}) , \qquad P_{B} = k/(2a^{2})\,.$$
The Friedmann equation is then simply:
\begin{align}
    3H^{2} = \rho - 3k /(2a^{2}).
\end{align}

Henceforth we shall refer to the collection of components which scale as $a^{-2}$ simply as ``curvature''. We see then that this contains a contribution from the spatial Ricci scalar $^{3}R$ (itself differing from the FRW case by a factor of 3) and the energy density of the field $B_{ab}$ (absent in the FRW case). The proportional contribution of `curvature' to the critical energy density is then given by
$$\,\Omega_{k0} \equiv -k/(2H_{0}^{2})\,$$
and so
$$\,|k|= 2H_{0}^{2}|\Omega_{k0}|\,.$$

We may note that the model may be written in terms of a vector field
$W^a$ constrained by a 2-form Lagrange multiplier $\,\lambda_{ab}\,$
as $$\,\mathcal{L} = W^{2}+ (\nabla^a W^b)\lambda_{ab}\,.$$

In contrast to several recent proposals employing higher spin fields in cosmology, our model avoids the generic instabilities which may ensue from fixed-norm constraints, potential terms or non-minimal couplings~\cite{Himmetoglu:2008zp}. In fact, as is well known, the canonical 1-form theory provides us electromagnetism and the canonical 3-form theory yields a contribution to the cosmological constant \cite{Hawking:1984hk} whose magnitude can then be fixed in relation to the observed value.

\section{Distances in LRS Metrics}

Let us uncover the observable implications of this set-up. The LRS class of metrics describe spacetimes with spatial sections orthogonal to the flow of physical time which contain both flat and curved two dimensional surfaces.

For instance consider the Kantowski-Sachs metric, for which the spatial sections are $\mathcal{R} \times \mathcal{S}^2$: any photon which traverses a distance in the $y - \xi$ plane is essentially propagating on the surface of a cylinder of time-dependent size (remember we are now assuming $a(t) = b(t)$). Because of this, the functional form of the angular diameter distance $d_{A}$ will be as in the flat FRW case ($d_{A}= a\chi$), however, the functions $a(t)$ and $\chi(a)$ will \emph{only} correspond to the actual flat FRW case if $S(\xi)=\xi$.

In LRS metrics, there are two distinct ways of defining $\,d_{A}$:
\begin{itemize}
    \item[(i)] as the angular diameter distance associated with an one-dimensional object of intrinsic (proper) length $L$ perpendicular to the line-of-sight and which subtends a (small) angle $\alpha$ in the sky as:
    $ d_{1A} \,\equiv\, L/\alpha\,$;
    \item[(ii)] as the solid-angle distance associated with a 2-dimensional object or intrinsic (proper) surface area $A$ which subtends a solid angle
    $\Omega$ as $ d_{2A} \,\equiv\, \sqrt{A/\Omega}$.
\end{itemize}
For a FRW metric, both definitions are equivalent. But that is not the case in LRS metrics. The second definition is the one appropriate to fit standard cosmological observations.

The reason is $d_{1A}$ depends not only on the position of the source in space, but also on its~\emph{orientation}. To visualize this one can consider two different slicings of the spacial sections of spacetime assuming $\,k\neq 0\,$ and the observer at the center of the
coordinate system. In the plane $\,y-\xi\,$ the metric is flat and photons propagate as in flat space. Therefore, a small one-dimensional line object aligned along the constant $\phi$ direction has a flat-like angular diameter distance. On the other hand, on the $\,\xi-\phi\,$ surface the space is curved (spherically or hyperbolically, depending on $k$), so if we put this rod in this plane the photons traveling to the center propagate according to a curved metric.
%

To write down the results explicitly, one rewrites the metric~\eqref{eq:lrs-metric} in standard spherical coordinates, using $\chi$ as radial coordinate and $\theta$ as zenith angle, the angle with respect to the preferred direction $V^{a}$:  
\begin{equation}\label{eq:ds2-th-phi}
    \dd s^2 = - \dd t^2 + a^2 \! \left[\dd \chi^2  + \chi^2  \dd \theta^2 + \frac{1}{|k|}S^2 \!\left[|k|^{\frac{1}{2}}\chi \sin\theta\right]\! \dd \phi^2 \right] \!.
\end{equation}
Consider an observer at a point ${\cal P}$ in the co-moving coordinates. It can be seen from the geodesic equations that all photons incident upon ${\cal P}$ at a given time follow trajectories of constant $\theta$ and $\phi$.


The solid angle subtended by a patch $\{\dd \theta \dd \phi\}$ is just given by $\,\sin\theta \dd\theta \dd\phi$, and its (proper) surface area by
$$\, \frac{a^{2}\chi}{\sqrt{|k|}}\, S\left(\chi \sqrt{|k|} \sin\theta\right) \dd \theta \dd \phi\,.$$
Therefore
\begin{equation}\label{eq:d2a}
    d_{2A}(\theta)^2 \,=\, \frac{a^2(t)\,\chi}{H_{0}\sqrt{2\,|\Omega_{k0}|}} \, \frac{ S\left(H_{0}\sqrt{2\,|\Omega_{k0}|} \,\chi \sin\theta\right)}{{\sin\theta}}\,,
\end{equation}
which is to be compared to the FRW equivalent:
\begin{equation}\label{eq:d2a-FRW}
    \left[d_{2A}^{\rm FRW}\right]^2 \,=\, \frac{a^2(t)\,}{H_0{}^2 \, |\Omega_{k0}|} \; S^2\left(H_{0}\sqrt{|\Omega_{k0}|} \,\chi \right)\,.
\end{equation}

Similarly, we obtain the  1-D angular diameter distance by considering a rod aligned so that its center is normal to the vector $(\partial_{\chi})^{a}$, subtending some angle $\alpha$ and spanning a distance $\,L/a$ in our comoving coordinates:
\begin{equation}\label{eq:d1a}
    d_{1A}(\theta, \omega)^2
    = a^2(t) \frac{\chi^2 \, \tan^2 \omega + \frac{S^2 \left(H_{0}\sqrt{2|\Omega_{k0}|}\,\chi  \sin\theta\right)}{H_{0}^2 2|\Omega_{k0}|}}{\tan^2\omega + \sin^2 \theta} \!
\end{equation}
where $\omega$ is the angle of orientation, defined as
$\,\tan \omega \equiv \delta \theta / \delta \phi$. From~\eqref{eq:d1a} one sees that the extreme values of $d_{1A}$ with respect to $\omega$ are just $\,d_{1A}^{\rm ext_1} = a(t) \chi\,$ and
$$\,d_{1A}^{\rm ext_2} = a(t) S(H_{0}\sqrt{2\,|\Omega_{k0}|} \chi) / H_{0}\sqrt{2\,|\Omega_{k0}|}\,.$$
It can be shown that $d_{2A}$ is an intermediate value between these two extrema, as one may have expected.

Observations that use the angular diameter distance, such as gravitational lensing, should be defined using 2-dimensional objects. Therefore henceforth we will always mean $d_{2A}$ whenever we mention angular diameter distances. The luminosity distance can then be obtained directly, as it is always related in the same way to the solid angular diameter distance ($d_{2A}$) due to the Reciprocity Theorem~\cite{1933PMag...15..761E}: in any Riemannian metric gravitational theory one has the relation $d_{L}=d_{2A}/a^{2}$.

\section{Observational Constraints}\label{sec:observ}

\subsection{CMB Constraints}\label{sec:obs-cmb}

We first consider whether the spacetimes here admit an isotropic
radiation field and that this field will be measured as isotropic by
the observers co-moving with the cosmological dust. The former
consideration may be addressed using the theorem of Ehlers, Geren, and
Sachs~\cite{Ehlers:1966ad} which states that there exist isotropic
solutions of the collisionless Boltzmann equation if the spacetime is
conformally equivalent to a stationary spacetime i.e. a spacetime with
a timelike Killing vector $\xi^{a}$ which is hypersurface orthogonal
(thus satisfying $0=\epsilon_{abcd}\xi^{a}\nabla^{b}\xi^{d}$).

Changing our time co-ordinate to $\eta=\int(1/a(t))dt$ (sometimes referred to as conformal time) we have that:
\begin{align}\label{eq:lrs-metric2}
    \!\!\!\dd s^2 = a(\eta)^{2}\left( - \dd \eta^2 \!+\!\dd y^2 \!+\!\! \dd \xi^2 \!+\! \frac{1}{|k|}S^2(|k|^{\frac{1}{2}}\xi) \dd \phi^2 \right)
\end{align}
By inspection the so-called conformal Killing field $(\partial_{\eta})^{c}$ is hypersurface orthogonal. The theorem states that it is the field $a(\partial_{\eta})^{c}$ with respect to which an isotropic radiation field may be measured. This indeed coincides with our assumed four-velocity of dust.

We can get a sense for how this would be realized in an experiment to observe the background CMB. Such experiments measure an amount of space volume from the (three-dimensional) surface of last scattering. Consider a function ${\cal N}(x^{i},p_{i})$, defined such that in the last scattering surface the number of photons between $\textbf{x}$ and $\textbf{x}+\textbf{dx}$ and with momentum between $\textbf{p}$ and $\textbf{p}+\textbf{dp}$ is given by ${\cal N}\sqrt{h}d^{3}x d^{3}p$ where $h_{ab}$ is the 3-metric on the last scattering surface.

The experiment, in operating for a time $\dd t$ will collect photons from the volume
$$\int_{\Omega} a^{3}_{*}\chi_{*}S(\chi_{*}\sin(\theta))\dd\chi \dd\Omega $$ where $a_{*}$ is the scale factor at last scattering and $\chi_{*}$ is the co-moving distance to last scattering at cosmic time $t_{0}$ (now).

As we have seen,  these background photons travel to us on geodesics $(\theta,\phi)=$ const $\,$ (in the coordinates of~\eqref{eq:ds2-th-phi}) so $a_{*}\dd\chi= \dd t$. Thus the number $\delta N$ of CMB photons received is:
\begin{eqnarray}
\label{eq:ps}
    \delta N &=& \int_{\Omega}\int_{\textbf{p}} a^{2}_{*} c\chi_{*}\frac{S(\chi_{*}\sin(\theta))}{\sin\theta}{\cal N}\dd t \dd\Omega \dd^{3}p\,.
\end{eqnarray}
The volume probed on the last scattering surface will be far smaller than the radius of curvature at recombination and so we should expect a-priori that ${\cal N}$ simply takes the familiar form
$${\cal N} = \frac{{\cal N}_{c}}{e^{\frac{p}{k T_{*}}}-1}$$
where ${\cal N}_{c}$ is a constant independent of $x^{i}$ and $p_{i}$ and $T_{*}$ is the temperature at last scattering.

We now must relate the amount of momentum space probed in terms scales typical of the experiment. Consider a point source radiating on the surface of last scattering and a CMB detector with cross sectional area A. The angular range $\Delta\beta$ of momenta probed from this point will be $\Delta\beta= A/d^{2}_{2A}$, where it has been assumed that $A$ is substantially smaller than $d^{2}_{2A}$, as expected. Thus, assuming a sufficiently small solid angle is probed, the experiment probes a momentum space volume of $\dd^{3}p=p^{2}\dd p\Delta\beta=p^{2}(A/d^{2}_{2A})\dd p$. We see then that the momentum space measure introduces an angular dependence which precisely cancels out that introduced by the space volume measure. One expects then to see an isotropic blackbody form for the CMB distribution function, precisely as in the FRW case.

Since at the background level the CMB remains isotropic, any constraints on anisotropic curvature from the CMB must come from perturbations, which have to be worked out and will appear in future work. Moreover, since the equations are much more complicated in the present model with anisotropic curvature than in standard cosmology, it is non-trivial to guess a priori what will be the dominant effects. In the absence of a better motivation, we might speculate that the constraints on curvature will be of the same order of magnitude, and below we review some of them.

The location $\ell^{(1)} $ of the first acoustic peak in the CMB
temperature anisotropy (in a FRW spacetime) is well approximated by 
$ \ell^{(1)} = \pi d_{2A}(z_*)/d_S(z_*)\,$, where $d_{2A}$ is the two dimensional angular diameter distance given by~(\ref{eq:d2a}) at the surface of last scattering (at redshift $z_{*}$) and $d_S(z_*)$ is the sound horizon at last scattering. When $|\Omega_{k0}|$ is small, Eq. (\ref{eq:d2a}) and some algebra lead to
\begin{align}
    |\Omega_{k0}|^{\rm FRW} = \frac{3}{(H_0\chi_{*})^2} \frac{\Delta\ell^{(1)}}{\ell^{(1)}}\,,
\end{align}
where we emphasize the fact that this result is of yet not known to be valid also in LRS spacetimes. The value of $\ell^{(1)} $ will depend on the orientation of the probed part of the sky with respect to the preferred direction. To estimate the bounds on anisotropy,  we assume that the background expansion is such that the all-sky average location of the peak is the observed one. This implies $H_0\chi_{*} \approx 7/2$, supporting our previous assumption. From~\cite{Eriksen:2004}, we see that the data allows at two (one) $\sigma$ level about twelve (six) percent variations in the peak location across the sky. These bounds translate to $|\Om_{k0}|^{\rm FRW}<0.03$ ($|\Om_{k0}|^{\rm FRW}<0.015$). In comparison, WMAP 5-year data together with $H_0$ measurements $-0.052 <
\Omega_{k0}{}^{\rm FRW} < 0.013$ at $2\sigma$ (assuming a \lcdm\ model -- bounds are a little looser if one allows $w \neq -1$)~\cite{Komatsu:2008hk}.

\begin{figure}[t]
    \hspace{-1cm} \includegraphics[width=7.2cm]{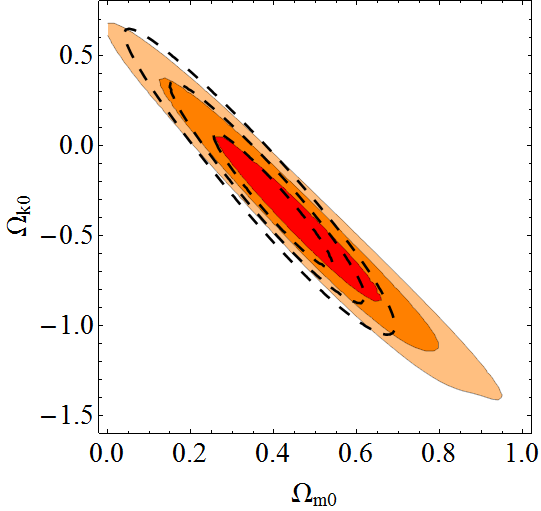}
    \vspace{.2cm}

    \includegraphics[width=8cm]{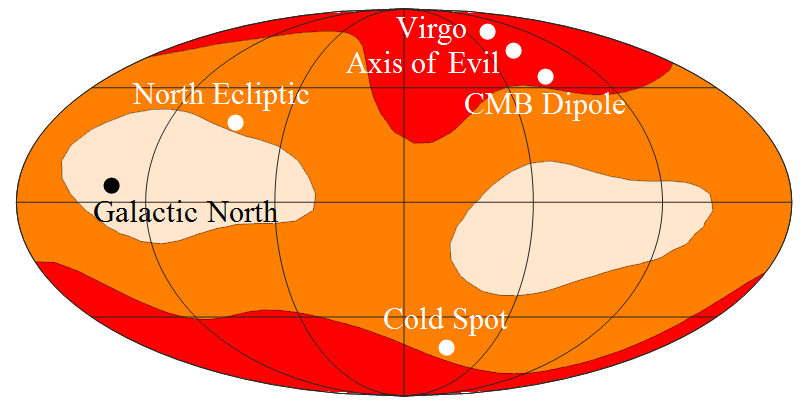}
    \caption{\label{contours} Top: 1, 2 and 3$\sigma$ confidence-level       contours obtained from the SN Ia data. Solid contours stand for the model considered here, thick dashed lines for $\Lambda$CDM (marginalized over $H_0$, $u_{P}$ and $v_{P}$). Bottom:  constraints on the preferred direction in galactic coordinates. Red (darker) and orange (medium) stand for 1 and 2$\sigma$, whereas the light yellow (lighter) stand for all regions excluded at more than 2$\sigma$ (marginalized over $H_0$, $\Omega_{m0}$ and $\Omega_{k0}$). Some positions in the sky are marked in the figure as reference.}
     \vspace{-.2cm}
\end{figure}

\subsection{Supernova Constraints}\label{sec:obs-sne}

We performed a likelihood analysis of the model comparing the predicted and the observed apparent luminosity of each SN in the Union sample~\cite{Kowalski:2008ez}. Here the prediction depends not only on the redshift but also on the angular position of the SNe.  If the preferred direction is at $(u_{P},v_{P})$ in the astronomers' angular coordinates  $(u,v)$ then we have that the `angle with respect to the preferred direction' $\theta=\theta(u_{P},v_{P},u,v)$ is given by
\begin{align}
    \nonumber\cos \theta =  \sin u\sin u_{P}\left[\cos (v-v_{P})\right] +\cos u\cos u_{P}\,.
\end{align}
%

SNe here are even more flexible with respect to the value of
$\Omega_{k0}$ than in $\Lambda$CDM. This is a result of two competing
effects:
\begin{itemize}
\item[(i)] the anisotropic curvature does not affect SNe in the
  preferred direction;
\item[(ii)] $\,|k|= m H_{0}^{2}|\Omega_{k0}|$, where $m=2$ here but $m=1$ in $\Lambda$CDM, so SNe perpendicular to the preferred direction feel more curvature. This is well illustrated in the top panel of Figure~\ref{contours}. In fact, the present SN data alone do not tell us much and allow anisotropy and curvature to be present quite abundantly.
However, this result gives us a hint on what to expect from a future
CMB analysis, where the bound on curvature should also be looser.
\end{itemize}
As can be seen in the Figure~\ref{contours} (bottom), supernovae exhibit
no strongly preferred direction in this model; all one gets from the data is two broad regions in the direction pointing away from the galactic plane which is preferred at the $1\sigma$ level. The reader should note that care must be taking when interpreting this plot. Contrary to the usual confidence contours in cosmology, here the domain of the (marginalized, 2-parameter) likelihood is finite, and therefore by definition no matter what the quality of the data, all $\sigma$ levels will be represented. Thus, there will be always at least one region disfavored at an arbitrary $\sigma$ level, but that will have no physical significance. The real physical significance will be given by the presence or not (and their size) of 2 small antipodal regions strongly favored against all others.

One should also keep in mind that in the foreseeable future we expect to have orders of magnitude more SNe measurements, and it is a possibility that eventually the most stringent constraints on this kind of anisotropy ensue from the observations of the late-universe phenomena like the SN light curves. With this in mind, in the next section we investigate in more detail the observability of such anisotropy with future supernova surveys.

\section{Forecast of Future Observability with Supernovae}

In the previous section we analyzed the current SN constraints assuming a our LRS model. Here, we want to forecast how would future surveys fare in this respect. Our motivation is twofold: (i) to determine how many supernovae would be necessary to pinpoint a possible preferred direction in the sky in an anisotropic curvature scenario; (ii) to better interpret the results of the previous section, in particular Figure~\ref{contours}.

First, note that in~\eqref{eq:d2a}, the argument of the $S$ function is of order ${\cal O}(\sqrt{\Omega_{k0}})$. In fact, for all values of $\Omega_{m0}$ and  $\Omega_{k0}$ inside the $3\sigma$ contours defined by SNe alone, one has that $H_0 \chi(z) < 2$ for $z<2$. This motivates an expansion in powers of $\Omega_{k0}$, especially since (as discussed in section~\ref{sec:obs-cmb}) other observations imposes (in a FRW spacetime) a limit $|\Omega_{k0}| \lesssim 0.03$ an we speculate that in a LRS spacetime this bound will be of the same order of magnitude.

Expanding both~\eqref{eq:d2a} and the FRW equivalent~\eqref{eq:d2a-FRW} in powers of $\Omega_{k0}$ one gets that the difference between the distance moduli in both LRS and FRW can be written as:
\begin{align}
    \Delta \mu \, & \equiv \,\mu_{\rm LRS} - \mu_{\rm FRW} \,=\, \nonumber \\
    &- \frac{5}{6\ln[10]}\, \Omega_{k0}\, H_0 \,\chi(z) \cos^2\theta + {\cal O} \left( \Omega_{k0}{}^2\right),
\end{align}
which is correct for both signs of $\Omega_{k0}$. In order to better evaluate the amplitude of this difference, we can make use of an approximation formula for $\chi(z)$. In the range $0<z<2$, $\,|\Omega_{k0}|<0.2\,$ and $\,0.2<\Omega_{m0}<0.4$, a good approximation (with discrepancy always smaller than 20\%) is given by:
\begin{align}
    \chi(z,\Omega_{m0},\Omega_{k0}) H_0 \approx \left(0.8 - 0.3 \Omega_{k0}\right) \left(0.95 z - 0.18 z^2 \right).
\end{align}
Therefore, a rough expectation for $\Delta \mu$ can be achieved by just the leading terms, which give
\begin{align}
    \Delta \mu \, \approx \, -0.28 \, z \, \Omega_{k0}\,\cos^2\theta \,.
\end{align}

\begin{figure}[t]
    \includegraphics[width=8cm]{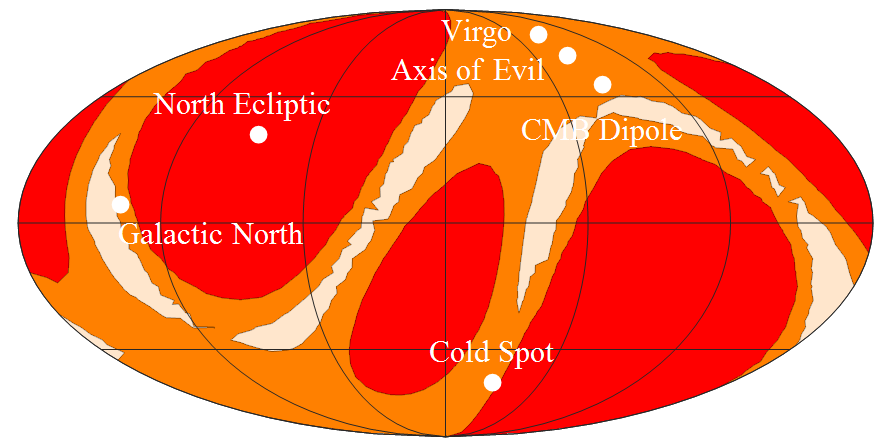}
    \includegraphics[width=8cm]{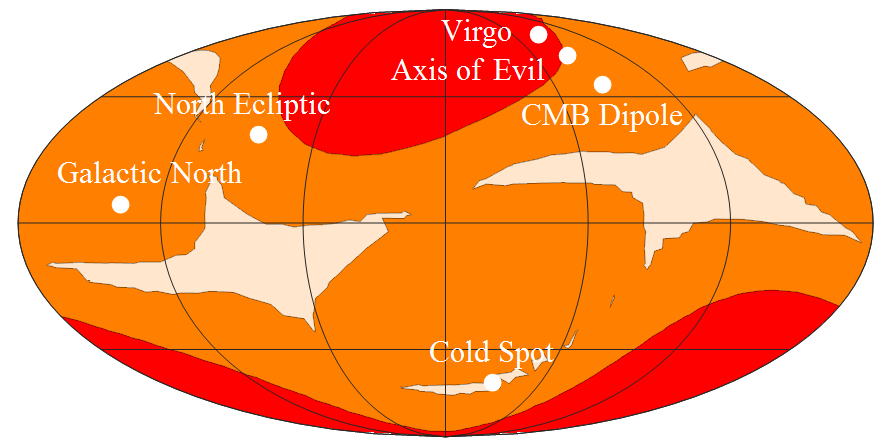}
    \includegraphics[width=8cm]{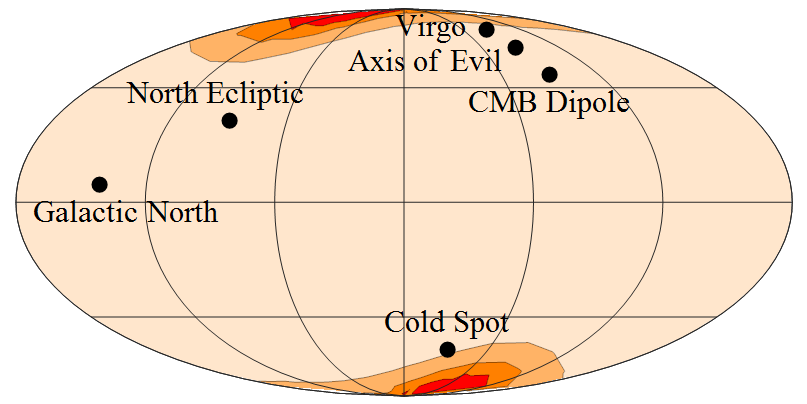}
    \caption{Constraints on the preferred direction in galactic coordinates for 3 supernova mock catalogs, assuming a LRS metric with $\Omega_{k0}=-0.1$ and preferred direction along the vertical axis as the fiducial model. Top: 1000 SNe; middle: 3000 SNe; bottom: 10000 SNe. Red (darker), orange (medium) and light orange (lighter -- bottom plot only) stand for 1, 2 and 3$\sigma$, respectively (marginalized over $H_0$, $\Omega_{m0}$ and $\Omega_{k0}$). The light yellow (lightest) represent the regions excluded at more than $2\sigma$ (above 2 plots) or $3\sigma$ (bottom).
    \label{fig:mock}}
     \vspace{-.2cm}
\end{figure}

To get a rough estimate on the number of supernovae required to spot a
spatial variation, it is interesting to compute the average
signal-to-noise ratio produced by a LRS metric in a supernova
catalog. The signal is just $ \Delta \mu$, whereas the noise is given
by the intrinsic dispersion of supernovae moduli, usually estimated to
be  around $\sigma_\mu \approx 0.15$mag (see below), which (unless limited by systematics) goes down as $N^{-1/2}$. The signal-to-noise ratio for $N_{\rm SNe}$ supernovae is thus
\begin{align}\label{eq:s-to-n}
    \frac{\rm Signal}{\rm Noise} \sim 0.28 \,\Omega_{k0} \, \langle z \rangle  \langle \cos^2\theta \rangle \frac{\sqrt{N_{\rm SNe}}}{0.15} \approx 0.65 \Omega_{k0} \sqrt{N_{\rm SNe}}\,,
\end{align}
where we assumed for the survey an average redshift $\langle z \rangle = 0.7$ and used the fact that $\langle \cos^2\theta \rangle = 1/2$.
Therefore, in order to obtain, say, a signal-to-noise ratio of 3, one needs \begin{align}\label{eq:nsne}
    N_{\rm SNe} \,\gtrsim \; \frac{20}{\Omega_{k0}{}^2}\,,
\end{align}
which for $|\Omega_{k0}| = 0.1$ corresponds to $N \sim 2000$ SNe. These rough estimates can be easily improved by specifying the expected number of supernovae per redshift bin of a given survey.

In order to get more precise results, we generated three supernovae
mock catalogs with $N_{\rm SNe} = 1000$, $3000$ and $10000$. We adopt
reasonable specifications for near-future surveys: all SNe were
assumed to be evenly distributed over the sky (which in practice
require either a space telescope or a combination of surveys in both
hemispheres of the planet), with a constant redshift distribution in
the range $\,0.1 <z< 1.2\,$ (for simplicity, although a more realistic
distribution will not alter much our results) and with an intrinsic
dispersion of $\,\sigma_\mu \approx 0.15$mag. The latter is defined as
the observed Hubble diagram scatter for ideal measurements, which is
around 0.15mag for modern surveys (but might be reduced to 0.10mag in
future experiments)~\cite{Astier:2010qf}.

As for the number of SNe, the Dark Energy Survey is expected to detect between $1000$ and $5000$ SNe (depending on the observing strategy) in a 5000${\rm deg}^2$~\cite{Bernstein:2009ue}; the Large Synoptic Survey Telescope (LSST) is expected to see at least $\sim 150000$ Type Ia SNe / year (during 10 years) in a 20000${\rm deg}^2$ area~\cite{LSST:2009pq}; Pan-STARRS is expected to detect $\sim 10^5$ SNe over a 30000${\rm deg}^2$ area~\cite{pan-starrs}.

A LRS metric with $\Omega_{k0} = -0.1$ and preferred direction along the Milky Way axis (vertical axis of Figure~\ref{contours}) was taken as the fiducial model. The results are depicted in Figure~\ref{fig:mock}. As expected by the above arguments, only for more than $\sim 2000$ SNe can one start to distinguish the preferred direction. For 1000 Sne, statistical noise is dominant. In particular, our rough estimate of the signal-to-noise ratio for the cases of Figure~\ref{fig:mock} yield S/N~=~$\{1.9, 3.3, 6.1\}$ for the \{top, middle, bottom\} plots.

In light of~\eqref{eq:nsne} and of Figure~\ref{fig:mock}, we are in a better position to interpret the results obtained with the Union SNe compilation. Inspection of Figure~\ref{contours} shows some similarities with the middle plot of Figure~\ref{fig:mock}, which assumes a somewhat high amount of curvature ($|\Omega_{k0}| = 0.1$), has 10 times the number of SNe, distributed over the whole sky and with smaller scatter ($\sigma_\mu^{\rm Union} \approx 0.28$, although in this catalog the reported value varies significantly, depending on the supernova). In fact, since the catalog contains only 307 SNe, according to~\eqref{eq:s-to-n} the expected signal-to-noise ratio is $S/N \sim 11 \,\Omega_{k0}$, so one should only expect to detect a possible preferred LRS direction (say, with $S/N \ge 3$) if $|\Omega_{k0}| \gtrsim \sqrt{20/307} = 0.26$, which is much higher than allowed by other observables such as a combination of CMB and $H_0$ measurements. This is a curious result, and may be just statistical fluke or indicate the presence of an unknown systematic in this sample (which indeed is a compilation of measurements done with a number of different instruments and surveys). We leave a more detailed investigation of this to a future work.

\section{Conclusions}

We considered the possibility of anisotropically curved
cosmology. This type of anisotropy has been heretofore largely
ignored, though shear-free LRS metrics were introduced as a
theoretical possibility almost twenty years ago~\cite{Mimoso:1993ym}
and a cosmological model utilizing them was explored almost a decade ago~\cite{Carneiro:2001fz}.

As cosmology strives to become a precision science, a complete understanding of all non-standard possibilities becomes imperative. This kind of anisotropy could be ``hidden'' within an effective FRW set-up, as the expansion of the background is precisely the same and in particular is isotropic, as is the CMB at the background level. It was shown that a canonical massless Kalb-Ramond field can support such a configuration and that the curvature anisotropy can in principle be detected either by CMB (at the perturbation level) or by future SN Ia data.

We showed that future SNe surveys could detect a modulation of distance moduli induced by LRS metrics by observing $\,\sim 20 / (\Omega_{k0}{}^2)$ supernovae over the whole sky, which indeed will be achieved by planned future surveys such as the LSST and Pan-STARRS as long as $|\Omega_{k0}|^{\rm LRS}$ is not much smaller than 0.01 (not to be confused with the bounds on $|\Omega_{k0}|^{\rm FRW}$ which are a priori different). Using the Union sample of SNe, a higher-than-expected angular dependence was derived, which warrants further investigation.

A full analysis of the proposed model (including the study of the perturbation equations) and of the observational consequences are under investigation and will be published in future work.

It remains to be seen whether the presence of such field can explain the anomalies in the CMB data~\cite{Copi:2010na}. Such a possibility would open a completely new window not only on the nature of the CMB anomalies but also into high energy physics beyond the Standard Model and the usual isotropic candidates of dark energy such as scalar fields or the cosmological constant.

\section{Acknowledgements}

We thank Marek Kowalski for SN data. We also thank Julian Adamek, Bruce Bassett, Jose Blanco-Pillado, Iain Brown, Saulo Carneiro, Hans Eriksen,  José  Mimoso, Ribamar R. R. Reis, Michael Salem, Mark Sullivan and Joe Zuntz for fruitful discussions. TSK is supported by Academy of Finland and the Yggdrasil grant of the Research Council of Norway. DFM thanks Research Council of Norway FRINAT grant 197251/V30.  MQ thanks UiO for hospitality during part of this project.

$\;$


\bibliography{sigrefslett}

\end{document}